\documentclass[%
 aip,
 amsmath,amssymb,
 reprint,%
]{revtex4-1}

\usepackage{graphicx}
\usepackage{dcolumn}
\usepackage{bm}

\usepackage[utf8]{inputenc}
\usepackage[T1]{fontenc}
\usepackage{mathptmx}
\usepackage{etoolbox}
\usepackage{bigints}
\usepackage[dvipsnames]{xcolor}
\usepackage{comment} 
\usepackage{ragged2e}

\makeatletter
\def\@email#1#2{%
 \endgroup
 \patchcmd{\titleblock@produce}
  {\frontmatter@RRAPformat}
  {\frontmatter@RRAPformat{\produce@RRAP{*#1\href{mailto:#2}{#2}}}\frontmatter@RRAPformat}
  {}{}
}%
\makeatother
\begin{document}

\preprint{AIP/123-QED}

\title[CO$_2$ Hydrate Phase Diagram from Computer Simulation]
{Unveiling the CO$_{2}$ Hydrate Phase Diagram from Computer Simulation: Locating the Hydrate–Liquid–Vapor Coexistence and its Upper Quadruple Point}


\author{Jesús Algaba}
\affiliation{Laboratorio de Simulaci\'on Molecular y Qu\'imica Computacional, CIQSO-Centro de Investigaci\'on en Qu\'imica Sostenible and Departamento de Ciencias Integradas, Universidad de Huelva, 21006 Huelva Spain}

\author{Samuel Blazquez}
\affiliation{Dpto. Qu\'{\i}mica F\'{\i}sica I, Fac. Ciencias Qu\'{\i}micas, Universidad Complutense de Madrid, 28040 Madrid, Spain}

\author{Cristóbal Romero-Guzmán}
\affiliation{Laboratorio de Simulaci\'on Molecular y Qu\'imica Computacional, CIQSO-Centro de Investigaci\'on en Qu\'imica Sostenible and Departamento de Ciencias Integradas, Universidad de Huelva, 21006 Huelva Spain}


\author{Carlos Vega}
\affiliation{Dpto. Qu\'{\i}mica F\'{\i}sica I, Fac. Ciencias Qu\'{\i}micas, Universidad Complutense de Madrid, 28040 Madrid, Spain}

\author{María M. Conde$^*$}
\affiliation{Departamento de Ingeniería Química Industrial y del Medio Ambiente, Escuela Técnica Superior de Ingenieros Industriales, Universidad Politécnica de Madrid,
28006, Madrid, Spain.
}

\author{Felipe J. Blas$^*$}
\affiliation{Laboratorio de Simulaci\'on Molecular y Qu\'imica Computacional, CIQSO-Centro de Investigaci\'on en Qu\'imica Sostenible and Departamento de Ciencias Integradas, Universidad de Huelva, 21006 Huelva Spain}

\begin{abstract}

Carbon dioxide (CO$_2$) hydrates hold promising applications in capturing and separating CO$_2$ for climate change mitigation.
Understanding their behavior at the molecular level is therefore essential, and computer simulations have become powerful tools for exploring their formation and stability, providing valuable insights into their underlying mechanisms.
In this work, we perform molecular dynamics simulations to compute the three-phase coexistence line involving the stability region where CO$_2$ is in the vapor phase: CO$_2$ hydrate - liquid water - vapor. This computation was previously inaccessible using the traditional three-phase direct coexistence technique. To achieve this, we employ a novel solubility-based method, which allows us to accurately evaluate the coexistence line. 
Our results exhibit excellent agreement with experimental data and, for the first time, accurately reproduce the hydrate–liquid–vapor equilibrium line of the CO$_2$–water phase diagram.
Finally, we have determined the upper quadruple point (Q$_2$) where the four phases, namely hydrate, liquid water, liquid CO$_2$, and vapor, coexist. 
Our pioneering result for the Q$_2$ value shows remarkable agreement with experimental observations, validating the accuracy of our findings and representing a significant milestone in the field of gas hydrate research.

\end{abstract}

\maketitle
$^*$Corresponding authors: felipe@uhu.es and maria.mconde@upm.es\\

%

\section{Introduction}

Clathrate hydrates, or simply hydrates, are expected to play a central role in the global transition from fossil fuels to renewable energy sources.~\cite{lee2001methane,ruppel2017interaction,Velus2014a,hassanpouryouzband2018co2,ma2016review,dashti2015recent,cannone2021review,duc2007co2,choi2022effective} These strategic materials are non-stochiometric crystalline inclusion compounds formed by hydrogen-bonded water molecules (host) capable of encapsulating molecules of interest (guest) under the appropriate thermodynamic conditions.~\cite{Sloan2008a,Ripmeester2022a} 
These compounds are of great interest because of their applications as a methane (CH$_4$) reservoir,~\cite{lee2001methane,ruppel2017interaction} as a safe option for hydrogen (H$_2$) storage,~\cite{Velus2014a,lee2005tuning,Lokshin2004a} as a clean alternative for recovering nitrogen (N$_2$) from industrial gas emissions,~\cite{Yi2019,hassanpouryouzband2018co2} and as a stable method for CO$_2$ capture~\cite{ma2016review,dashti2015recent,cannone2021review,duc2007co2,choi2022effective,lee2014quantitative} among others. The case of CO$_2$ hydrates is particularly relevant because of their significance in the context of global climate change.~\cite{Sloan2008a,Ripmeester2022a} In recent years, considerable effort has been devoted to developing viable strategies for the secure capture and storage of CO$_2$ to mitigate this pressing challenge, and the use of hydrates as molecular containers for CO$_2$ represents a promising, albeit technically demanding, approach. For these reasons, CO$_2$ hydrates have been extensively investigated through both experimental studies~\cite{Servio2001a,sum1997measurement,makiya2010synthesis,choi2022effective,lee2014quantitative,cordeiro2019phase,koh2012recovery,pan2020new,daniel2015hydrate} and molecular simulations.~\cite{Miguez2015a,Perez-Rodriguez2017a,Waage2017a,Constandy2015a,Zhang2019a,Algaba2023a,Pineda2023a,Barnes2013a,phan2021molecular,wang2021promotion,jacobson2010nucleation,jacobson2010amorphous,Roman2010a,Zeron2025a,Romero-Guzman2023a,Algaba2024a,Algaba2024b}

Accurate knowledge of the thermodynamic stability of gas hydrates, particularly their phase diagrams, is crucial for their efficient utilization in energy production, carbon sequestration, gas storage, and transportation.~\cite{Sloan2008a,Ripmeester2022a} Phase diagrams delineate the conditions under which hydrates form and remain stable, thereby enabling the identification of optimal and cost-effective conditions for clathrate hydrate formation and application. Over the past several decades, extensive experimental studies have determined the hydrate dissociation line, which typically represents a three-phase equilibrium in a binary system, where hydrate, aqueous, and guest-rich gas or liquid phases coexist, depending on the identity of the guest molecule. Authoritative treatments of this subject can be found in the monograph by Sloan and Koh~\cite{Sloan2008a} and in the recent volume by Ripmeester and Alavi~\cite{Ripmeester2022a}, which provides a comprehensive review of hydrate phase behavior.

Molecular simulation offers a complementary route for determining hydrate dissociation lines.~\cite{Allen2017a,Frenkel2002a,Sloan2008a,Ripmeester2022a} In particular, the direct coexistence (DC) method introduced by Ladd and Woodcock~\cite{Ladd1977a,Ladd1978a} was successfully adapted to hydrate systems by some of us,~\cite{Conde2010a,Conde2013a} who first applied it to CH$_{4}$ hydrates. Since the initial work of Conde and Vega,~\cite{Conde2010a} this approach has inspired numerous studies that have employed the DC method to determine the dissociation lines of various hydrates.~\cite{Conde2010a,Jensen2010a,Miguez2015a,Michalis2015a,Constandy2015a,Perez-Rodriguez2017a,Fernandez-Fernandez2019a,Fernandez-Fernandez2021a,Fernandez-Fernandez2022a,Blazquez2023b,Algaba2024a,Algaba2024b,Borrero2025a,Torrejon2025a,Blazquez2024a,Algaba2024c}

In a subsequent study, Tanaka and co-workers~\cite{Tanaka2018a} determined the mutual solubilities of CH$_{4}$ in water under conditions where the aqueous solution coexists with both the CH$_{4}$ hydrate and the pure CH$_{4}$ phase. The solubilities were estimated from the excess chemical potential of the solute in the aqueous phase, assuming infinite dilution. This theoretical framework is among the standard approaches employed within classical equations of state to locate three-phase coexistence conditions involving gas and/or solid phases.~\cite{Blas2000b,Miguez2011a,Miguez2015b} More recently, we have extended this methodology by computing solubilities directly from molecular simulations, thereby demonstrating that the so-called solubility method constitutes a distinct yet fully equivalent alternative to the DC approach for determining the CH$_{4}$ hydrate dissociation line.~\cite{Grabowska2022a} This technique has later been applied by several authors to predict dissociation conditions,~\cite{Algaba2023a,Algaba2023b,Torrejon2024b,Torrejon2025a} including CO$_2$ hydrates.~\cite{Algaba2023a} In accordance with this innovative method, previously used to account for the hydrate-water-CO$_{2}$ three-phase line, the solubility of CO$_2$ when an aqueous phase (L$_{\text{H}_2\text{O}}$) is in contact with a vapor phase (L$_{\text{H}_2\text{O}}$-V) and when in contact with a hydrate phase (H-L$_{\text{H}_2\text{O}}$) is calculated keeping the pressure constant and exploring different temperatures. At the three-phase coexistence conditions, the solubility obtained from both equilibria (L$_{\text{H}_2\text{O}}$-V and H-L$_{\text{H}_2\text{O}}$) have to be the same, or in other words, the temperature at which the solubility curves intersect at constant pressure is the three-phase equilibrium temperature ($T_3$). The employed method has been schematized in Fig.~\ref{esquema}, where we show snapshots of the
H-L$_{\text{H}_2\text{O}}$ (top left)  and L$_{\text{H}_2\text{O}}$-V (top right) coexistences and their corresponding solubility curves (blue and red, respectively). As can be seen in the scheme, and as is well known, the solubility of a gas in water increases as the temperature decreases; however, in the case of hydrate–water systems, the behavior is the opposite. The intersection between those curves is the triple point temperature
where the three phases H-L$_{\text{H}_2\text{O}}$-V coexist (a snapshot of the three phases in coexistence is shown in Fig.~\ref{esquema} bottom).
Indeed, the introduction of this novel approach and the results presented in the subsequent paragraphs signify a remarkable breakthrough in the field of research.

It is important to recall that Tanaka and collaborators~\cite{Tanaka2024a,Tanaka2025a} have recently determined the H–L$_{\text{H}_{2}\text{O}}$-V coexistence curve of the CO$_{2}$ hydrate by evaluating the chemical potentials of water and CO$_2$ along the hydrate–CO$_2$ and hydrate–water equilibrium boundaries, using the van der Waals–Platteeuw theory.~\cite{Platteeuw1957a,Platteeuw1959a} The intersection of these two lines defines the $T_3$ triple point. Although he also computes the hydrate–water and water–CO$_2$ solubilities, which could in principle be used to locate $T_3$, he chose to determine it through the chemical potential route. It should be noted that this approach is not a purely molecular simulation prediction, as it relies on the van der Waals–Platteeuw theoretical framework.~\cite{Platteeuw1957a,Platteeuw1959a}

\begin{figure}
\includegraphics[width=\columnwidth]{scheme.pdf}

\caption{Top left: Snapshot of the hydrate-liquid coexistence. Top right: Snapshot of the liquid-vapor coexistence. Middle: Schematic solubility of a gas as a function of temperature of an aqueous phase when in contact via a planar interface with a hydrate phase (left), vapor phase (right), and in a triple coexistence (center). Bottom: Snapshot of the triple coexistence hydrate-liquid-vapor. } 
\label{esquema}
\end{figure}

Despite the recent contributions by Tanaka and co-workers and by other authors,~\cite{Tanaka2024a,Tanaka2025a,Grabowska2022a,Algaba2023a,Algaba2023b,Torrejon2024b,Torrejon2025a} a critical gap persists in simulation-based studies: the accurate prediction of the hydrate–water–vapor coexistence line for systems containing guest molecules that can exist in both vapor and liquid phases, obtained solely from computer simulation without invoking any theoretical framework. To the best of our knowledge, all the simulation studies about the CO$_2$ hydrate, have been performed at pressures above $44.99\operatorname{bar}$, at which the CO$_2$ + water binary mixture exhibits a H-L$_{\text{H}_2\text{O}}$-L$_{\text{CO}_2}$ three-phase coexistence line where the hydrate, aqueous and liquid CO$_2$ phases coexist. At pressures below $44.99\operatorname{bar}$, the liquid CO$_2$ phase becomes a vapor phase (V) and the system shows a H-L$_{\text{H}_2\text{O}}$-V three-phase equilibrium. Both three-phase coexistence lines (H-L$_{\text{H}_2\text{O}}$-V and H-L$_{\text{H}_2\text{O}}$-L$_{\text{CO}_2}$) meet at a Q$_2$ quadrupole point\cite{Sloan2008a} at $283\operatorname{K}$ and $44.99\operatorname{bar}$. Although several simulation works have been dedicated to the study of the CO$_2$ hydrate phase diagram,~\cite{Miguez2015a,Perez-Rodriguez2017a,Constandy2015a,Algaba2023a} the study of the H-L$_{\text{H}_2\text{O}}$-V branched remains elusive.

From a molecular dynamics (MD) perspective, the problem in simulating the hydrate-water-vapor coexistence line arises from the inherently different nature of the phases involved. While the water and hydrate phases have comparable densities and compressibilities, allowing them to be modeled effectively under constant pressure ($NPT$) conditions, the vapor phase presents a much lower density and significantly higher compressibility. 
Particularly, the density of the CO$_2$ vapor phase is about one order of magnitude lower, and the compressibility is about an order of magnitude higher. That means that usual algorithms fail to keep the pressure constant when a condensed and a vapor phase coexist in the same simulation box. This difference causes barostats in MD simulations to struggle with maintaining stable pressure, especially when all three phases are present in the simulation box.~\cite{Marchio2018a} In other words, usual algorithms fail to keep the pressure constant when a condensed and a vapor phase coexist in the same simulation box. At this point, it is important to remark that both methods, DC and the original solubility method, present the same problem due to the vapor nature of the CO$_2$ phase. In fact, when the DC technique is applied to systems containing vapor phases, the barostat often struggles to equilibrate properly and to yield accurate pressure values.

When the DC technique is employed to determine the H-L$_{\text{H}_2\text{O}}$-V equilibrium line at low pressures, below the Q$_{2}$ of the CO$_{2}$ hydrate, the method artificially extends the H-L$_{\text{H}_2\text{O}}$-L$_{\text{CO}_2}$ three-phase line. This occurs because it fails to correctly reproduce the vapor phase—the system pressure does not equilibrate properly—thereby pushing the coexistence line into a metastable region and yielding an incorrect slope in the guest vapor region.

The topology of the phase diagram exhibited by CO$_{2}$ hydrate is not unique. Many other clathrate hydrates display similar features, whereas some systems exhibit distinctly different topologies. For hydrates formed from a single guest species, two general types of phase diagrams can be identified depending on the critical temperature ($T_{c}$) of the guest. When $T_{c}$ lies near or below the triple-point temperature of pure water, the phase diagram differs topologically from that of guests whose $T_{c}$ exceeds the water triple point.~\cite{Sloan2008a}  

For guests with relatively low $T_{c}$, the region of interest for hydrate formation does not involve a liquid phase of the guest, whose behavior remains effectively gas-like. Under these conditions, only the hydrate–liquid–vapor (H–L$_{\text{H}_2\text{O}}$–V) three-phase line exists. This line extends from higher temperatures and pressures toward lower values and terminates at a quadruple point (commonly denoted Q$_{1}$ in the literature), where several three-phase lines intersect, all involving ice I$_h$ as one of the coexisting phases.~\cite{Sloan2008a} This behavior is characteristic of CH$_{4}$ and N$_{2}$ hydrates. The less dense phase is conventionally labeled as V (vapor), although it is, strictly speaking, a supercritical fluid, since the conditions relevant to hydrate formation typically lie above the guest’s critical point. For CH$_{4}$ and N$_{2}$ hydrates, the density of the gas-like phase is sufficiently high that it is more appropriately described as a fluid rather than a true vapor. In particular, CH$_{4}$ hydrate does not exhibit the equilibration issues discussed above, since the CH$_{4}$-rich fluid under H–L$_{\text{H}_2\text{O}}$–V coexistence conditions possesses a mass density approximately one-half to one-third that of pure water. Under these conditions, the barostat employed in the DC technique performs reliably.  

In contrast, guests whose $T_{c}$ exceeds the triple point of water exhibit phase diagrams with a different topology. In the hydrate-forming region, the guest condenses to form a distinct liquid (L) phase. In this case, in addition to the H–L$_{\text{H}_2\text{O}}$–V three-phase line, a further hydrate–liquid–liquid (H–L$_{\text{H}_2\text{O}}$–L$_{\text{guest}}$) three-phase line appears. Here L$_{\text{guest}}$ denotes a guest-rich liquid phase (L$_{\text{CO}_2}$ in the case of the CO$_{2}$ hydrate). Systems of this type are hereafter referred to as liquid-like. The H–L$_{\text{H}_2\text{O}}$–L$_{\text{guest}}$ line extends from higher temperatures and pressures toward lower values and terminates at an upper quadruple point (denoted Q$_{2}$), where hydrate, water-rich liquid, guest-rich liquid, and vapor coexist. From this Q$_{2}$ point, the H–L$_{\text{H}_2\text{O}}$–V line continues toward lower temperatures and pressures until it meets the Q$_{1}$ quadruple point.~\cite{Sloan2008a} Such liquid-like behavior is typical of ethane, H$_{2}$S, CO$_{2}$, among others.~\cite{Sloan2008a} As previously discussed for CO$_{2}$ hydrate, the H–L$_{\text{H}_2\text{O}}$–V three-phase line cannot be accurately determined using the standard DC technique, as the vapor phase is not dense enough compared to the water-rich liquid phase for the barostat to operate properly.

In this work, we demonstrate that an extension of the recently proposed solubility-based method enables accurate calculation of the three-phase coexistence equilibrium in CO$_2$ hydrate systems. Using this innovative approach, we report—for the first time—the H-L$_{\text{H}_2\text{O}}$-V three-phase equilibrium (hydrate--water--CO$_{2}$ vapor) and the upper Q$_2$ quadruple point of the CO$_2$–water phase diagram, where four distinct phases coexist: CO$_2$ hydrate, liquid water, liquid CO$_2$, and vapor. Among the various hydrates that exhibit gas- and liquid-like behavior, we select CO$_2$-hydrate since it is one of the most extensively studied.~\cite{Sloan2008a,Ripmeester2022a} Its relevance to carbon capture and sequestration has led to a wealth of experimental data, making it an ideal benchmark system for testing the accuracy of molecular dynamics simulations.

The remainder of this paper is organized as follows. Section~II presents the molecular models and simulation details. The results and their discussion are given in Section~III, and the conclusions are summarized in Section~IV.

\section{Molecular models and simulation details}

In this work, water and CO$_2$ molecules are represented using the widely adopted TIP4P/Ice~\cite{Abascal2005b} and TraPPE~\cite{Potoff2001a} force fields, respectively. This combination has been previously employed in the literature to determine the three-phase coexistence curve of CO$_2$ hydrates.~\cite{Miguez2015a,Constandy2015a,Algaba2024a,Algaba2024b,Algaba2023a} Míguez \emph{et al.}~\cite{Miguez2015a} proposed a modification of the unlike water--CO$_2$ interactions by introducing a scaling factor $\xi = 1.13$, such
that $\epsilon_{\text{W--CO}_2} = 1.13 \left( \epsilon_{\text{W--W}} \, \epsilon_{\text{CO}_2\text{--CO}_2} \right)^{1/2}$, where $\epsilon_{\text{W--W}}$ and $\epsilon_{\text{CO}_2\text{--CO}_2}$ are the water--water and CO$_{2}$--CO$_{2}$ Lennard-Jones dispersive interactions. Notice that the factor $1.13$ is applied to the interaction of water with both the C and O atoms of the CO$_{2}$ molecule. This adjusted combination rule yields an accurate description of the CO$_2$ hydrate dissociation temperature along the entire three-phase coexistence line.~\cite{Miguez2015a,Algaba2023a,Algaba2024a,Algaba2024b}

Unfortunately, although this combination of the TIP4P/Ice water model, the CO$_2$ model, and the selected $\xi$ value performs well for the H–L$_{\text{H}_2\text{O}}$–L$_{\text{CO}_2}$ three-phase equilibrium, it does not reproduce CO$_2$ solubilities with the same accuracy. The TIP4P/Ice model with $\xi = 1.13$ systematically overestimates experimental CO$_2$ solubilities.~\cite{Miguez2015a} However, as will be explained later, our aim is not to determine solubility values themselves, but rather to identify the $T_3$ point, which corresponds to the intersection of the two solubility curves.

All MD simulations have been performed using GROMACS (2021.5).~\cite{VanDerSpoel2005a} In all cases, the Verlet-leapfrog algorithm\cite{Cuendet2007a} with a time step of $2\,\text{fs}$ was used for solving Newton's dynamic equations, as well as a cut-off, for the dispersive and coulombic interactions, of $1.0\,\text{nm}$. Following our previous works,~\cite{Miguez2015a,Algaba2023a,Algaba2024a,Algaba2024b} Berthelot combining rule has been modified for the water--CO$_2$ interactions as
$\epsilon_{\text{W--CO}_2} = \xi \left( \epsilon_{\text{W--W}} \, \epsilon_{\text{CO}_2\text{--CO}_2} \right)^{1/2}$, with $\xi=1.13$.  Also, PME\cite{Essmann1995a,Lundberg2016a} (particle mesh Ewald) long-range corrections were applied for the coulombic interactions. We have employed the Nose-Hoover thermostat\cite{Nose1984a} algorithm with a coupling time constant of $2\,\text{ps}$ to fix the temperature. In the case of the $NPT$ simulations, an anisotropic Parrinello-Rahman barostat\cite{Parrinello1981a} with a coupling time constant of $2\,\text{ps}$ was used to fix the pressure and avoid stress from the solid hydrate structure. 

 \section{Results}

 Following the solubility method,~\cite{Grabowska2022a,Algaba2023a,Tanaka2018a,Torrejon2024b,Algaba2023b,Torrejon2025a} previously explained, the three-phase coexistence equilibrium problem can be split into two coexistence equilibria of two phases (V-L$_{\text{H}_2\text{O}}$ and H-L$_{\text{H}_2\text{O}}$).  We first focus on the case of V-L$_{\text{H}_2\text{O}}$ equilibria, in which the pressure cannot be kept constant due to the difference in density and compressibility of the V and the L$_{\text{H}_2\text{O}}$ phases. Since the pressure cannot be fixed to a desired constant value, the $NVT$ canonical ensemble is the most appropriate ensemble for a system with a vapor-liquid (V-L$_{\text{H}_2\text{O}}$) interface.~\cite{Frenkel2002a,Allen2017a} Although the pressure cannot be fixed, there is still a method to get the solubility at the desired temperature and pressure conditions.  V-L$_{\text{H}_2\text{O}}$ $NVT$ simulations of $400\operatorname{ns}$ ($100\operatorname{ns}$ for the equilibration and $300\operatorname{ns}$ for the production period) can be carried out at constant temperature but varying the volume and/or the number of molecules in the simulation box. We have fixed the number of water molecules at $4000$ in all cases, but the number of molecules of CO$_2$ has been fixed at $300$ and $600$. The top-right snapshot of Fig.~\ref{esquema} depicts the L$_{\text{H}_2\text{O}}$--V simulation setup, with the water-rich liquid phase (L$_{\text{H}_2\text{O}}$) located on the left side of the simulation box and a planar interface separating it from the vapor phase (V) on the right. The $z$ side of the simulation box ($L_z$), which is the direction along the V-L interface is formed, has been extended from $40$ to $180\operatorname{nm}$ ($L_x$ and $L_y$ have a fixed value of $2.8\operatorname{nm}$). Keeping the temperature constant at $270$, $275$, $280$, and $285\operatorname{K}$, the different combinations of $L_z$ values and CO$_2$ molecules--see Table \ref{P-size}--allow to explore the V-L$_{\text{H}_2\text{O}}$ system at pressures ranging from $5$ to $50\operatorname{bar}$. 
 
 \begin{table}
\caption{Values of pressures calculated from the $NVT$ simulations as a function of the number of molecules of CO$_2$ ($N_{\text{CO}_2}$) and the dimension of the $z$ side ($L_z$) of the simulation box.}
\label{P-size}
\centering
\begin{tabular}{lccccc}
\hline\hline
$T$ (K) & &270  & 275  & 280  &  285   \\
\hline
$L_z$ (nm) & $N_{\text{CO}_2}$ &\multicolumn{4}{c}{$P$ (bar)} \\
\hline
180 & 300  &  6.76(1)  & 7.11(1)   & 7.57(1)  &   7.85(1)    \\
120 & 300  & 8.79(1)   & 9.41(2)   & 9.61(1)   & 10.52(1)    \\
180 & 600  & 14.66(2)  & 15.47(2)   & 15.69(2)   & 16.82(2)    \\
120 & 600  & 18.11(3)  &  19.20(3)  &  20.08(3)  &  20.83(3)     \\
80 & 600  &  23.68(3)  & 25.16(3)  & 26.42(4)   &  27.87(3)   \\
40 & 600  & 37.42(4)   & 40.49(5)   &  44.36(5)  &  47.91(4)   \\
\hline
\end{tabular}
\end{table}

\begin{figure}
\includegraphics[width=\columnwidth]{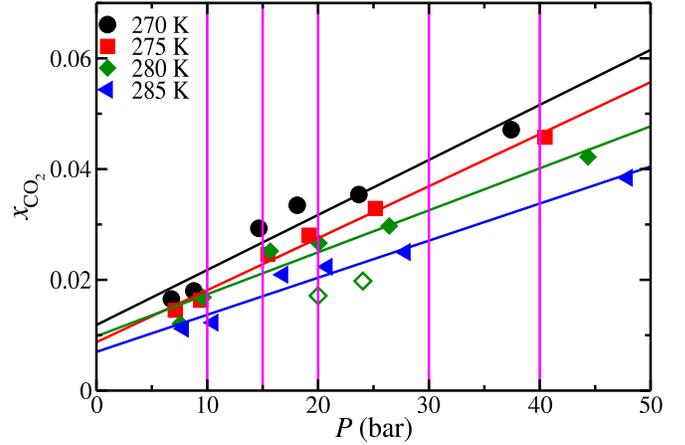}

\caption{Solubility of CO$_2$, as a function of pressure, for the four isotherms studied in this work in the aqueous phase when in contact via a planar interface with the vapor phase. The meaning of the symbols is explained in the legend. Magenta vertical lines represent the pressures chosen to interpolate the solubilities in this work. The continuous black, red, green, and blue lines have been obtained by linearly fitting the solubility results at each temperature. The cross points between the magenta lines and each fitting line correspond to the solubility values used for the $T_3$ determination. Green open diamonds represent the experimental solubility of CO$_{2}$ taken from the literature~\cite{Servio2001a,Wang2025a} at $280\operatorname{K}$ and two different pressures.}
\label{Solubility-VL}
\end{figure}

 The solubility results of each isotherm are obtained by analyzing the density profiles of each simulation and averaging the densities of water and CO$_2$ in the aqueous phase (note that density profiles have been obtained by dividing the simulation box into 200--1000 slabs depending on the size of the simulation box). In order to calculate the solubility of CO$_2$ at the desired pressure value, it is necessary to fit the solubility results at each isotherm as a function of pressure, as we show in Fig. \ref{Solubility-VL}. 
 From the graph, various conclusions can be drawn. Firstly, and as expected from Henry's law, the solubility of CO$_2$ increases with increasing pressure. Additionally, it is evident that lower temperatures result in higher solubility values (as expected in the case of gases, where solubility increases as temperature decreases). 
The obtained results align remarkably well with a linear fit, allowing us to accurately determine the solubilities of CO$_2$ at the desired pressure (marked as vertical pink lines) for each of the studied isotherms. Unfortunately, the CO$_2$ solubility in the aqueous phase in contact with the vapor phase is slightly overestimated when compared with experimental data reported in the literature.~\cite{Servio2001a,Wang2025a} In particular, the simulated molar fraction is approximately 0.01 higher than the experimental value at $280\operatorname{K}$, as shown in Fig.~\ref{Solubility-VL}. This deviation arises from the use of the TIP4P/Ice water model and the chosen value of $\xi$, as previously discussed in Section II.
 
\begin{figure}
\includegraphics[width=\columnwidth]{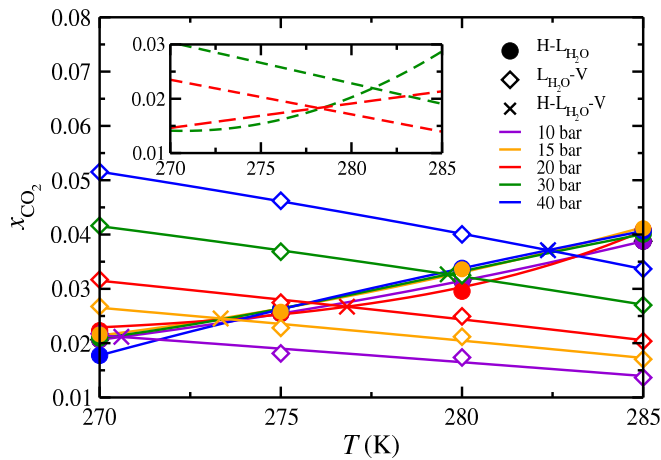}

\caption{Solubility of CO$_2$, as a function of temperature, at five pressures (10, 15, 20, 30, and $40\,\text{bar}$) of an aqueous phase when in contact via a planar interface with the vapor phase or the hydrate phase. The meaning of the symbols and lines in the main plot is represented in the legend. The red and green dashed curves shown in the inset correspond to correlations based on experimental data reported in the literature~\cite{Servio2001a,Wang2025a} for the solubilities of CO$_2$ in the aqueous solution with in contact with the vapor and hydrate phase along the $20$ and $30,\text{bar}$ isobars.} 
\label{Solubility-T3}
\end{figure}

We now study the case of the H-L$_{\text{H}_2\text{O}}$ equilibrium, since both are condensed phases, $NPT$ isobaric-isothermal simulations of $1200\,\text{ns}$ have been carried out, where the first $400\,\text{ns}$ are taken as the equilibration period and the last $800\,\text{ns}$ as the production period--note that in this case longer simulation times are required than in the V-L$_{\text{H}_2\text{O}}$ $NVT$ simulations because the dynamic of the solid hydrate phase is slower than that of the fluid phases.  The H-L$_{\text{H}_2\text{O}}$ simulation box consists of a hydrate composed of 4$\times$4$\times$4 unit cells, corresponding to $2944$ H$_2$O molecules and $512$ CO$_2$ molecules, in contact with an aqueous phase containing $4000$ H$_2$O molecules and $120$ CO$_2$ molecules. A schematic view of the H--L$_{\text{H}_2\text{O}}$ simulation setup is shown in the top-left snapshot of Fig.~\ref{esquema}, where the hydrate phase (H), located on the left side of the simulation box, coexists across a planar interface with the water-rich liquid phase (L$_{\text{H}_2\text{O}}$) on the right. This system size, for both the hydrate and the aqueous phase, is sufficient to avoid finite-size effects and is consistent with the cutoff radius employed.~\cite{Blazquez2024a,Algaba2024a,Algaba2024b}
The temperature as well as the pressure can be kept constant by using the classical combination of a barostat plus a thermostat algorithm.
The results obtained from the  H-L$_{\text{H}_2\text{O}}$ $NPT$ simulations are at the desired values of temperature (270, 275, 280, and $285\,\text{K}$) and pressure (10, 15, 20, 30, and $40\,\text{bar}$), and the density profiles can be used straightforwardly for solubility determination. In Fig. \ref{Solubility-T3}, we present first the solubility curves of
H-L$_{\text{H}_2\text{O}}$ (filled circles) at the five studied pressures. 
Interestingly, we observe that, similarly to the experiments\cite{Servio2001a}, at low temperatures, the highest pressure does not exhibit the highest solubility. Therefore, our results clearly agree with the experimental data.
If we now plot the solubility curve V-L$_{\text{H}_2\text{O}}$ as a function of temperature in Fig. \ref{Solubility-T3}, using the interpolations from the fits shown in Fig. \ref{Solubility-VL} as previously described, we can determine the $T_3$ values for the five isobars under consideration. At the three-phase coexistence conditions, the solubility obtained from both two-phase equilibria curves (V-L$_{\text{H}_2\text{O}}$ and H-L$_{\text{H}_2\text{O}}$) should be the same. Thus, we obtain the $T_3$ as the temperature at which the solubility isobars intersect each other on the graph. These intersection points are represented as crosses, each corresponding to the specific color of the respective isobar. As in the case of the aqueous phase in contact with the vapor phase, the CO$_2$ solubilities obtained for the aqueous solution when it is in contact with either the vapor or the hydrate phase are also slightly overestimated relative to experimental data reported in the literature.~\cite{Servio2001a,Wang2025a} In particular, the simulated molar fraction is about 0.01 higher than the experimental value at $280\,\text{K}$, as shown in Fig.~\ref{Solubility-T3}. This deviation likewise stems from the use of the TIP4P/Ice water model and the chosen value of $\xi$, as discussed in Section II.

A separate issue concerns the level of agreement between our model predictions and experimental data from the literature. It is important to note that the combination of water and CO$_2$ models employed here, together with the chosen value of $\xi$, has been used previously by some of us to accurately predict the hydrate–liquid–liquid three-phase equilibrium of CO$_2$ hydrate at higher pressures.~\cite{Miguez2015a} Unfortunately, although this combination of the TIP4P/Ice water model, the CO$_2$ model, and the selected $\xi$ value performs well for the H–L$_{\text{H}_2\text{O}}$–L$_{\text{CO}_2}$ three-phase equilibrium, it does not reproduce CO$_2$ solubilities with the same accuracy. The TIP4P/Ice model with $\xi = 1.13$ systematically overestimates experimental CO$_2$ solubilities.  This deviation is not solely attributable to the value of $\xi$, but is primarily due to intrinsic limitations of the water model. As shown in a recent study by some of us,~\cite{Blazquez2024c} the overestimation originates from discrepancies between the TIP4P/Ice and experimental densities of pure water below $300\operatorname{K}$. The physical origin of this overestimation lies in the water model. TIP4P/Ice predicts a liquid-water density that is systematically lower than the experimental one because it places the temperature of maximum density (TMD) at about $295\operatorname{K}$ instead of the experimental $277\operatorname{K}$.~\cite{Blazquez2024c} As a result, the model underestimates water density below $295\operatorname{K}$, and this deviation persists over the pressure range relevant to the gas branch ($1-50\operatorname{bar}$). A reduced liquid density makes it easier for the aqueous phase to accommodate CO$_2$ molecules, lowering their chemical potential and thereby increasing the predicted solubility. The $T_3$ point corresponds to the temperature at which the solubility of CO$_2$ in the aqueous solution when it is in contact with the vapor phase becomes equal to that of CO$_2$ when it is in contact with the hydrate phase. Although our model overestimates the solubility of CO$_2$ in water when it is in equilibrium with the gas phase, it also overestimates the solubility when the aqueous phase is in equilibrium with the hydrate. As a result, the two curves intersect at a temperature close to the experimental $T_3$.

Indeed, once the three-phase equilibria temperatures at different pressures have been calculated, we can encounter the challenge of constructing the phase diagram for pressures below $45\,\text{bar}$, where no previous equilibrium points were computed.
The H-L$_{\text{H}_2\text{O}}$-L$_{\text{CO}_{2}}$ line had been previously determined in several simulation studies\cite{Miguez2015a,Constandy2015a,Algaba2023a,Algaba2024a,Algaba2024b} but with this work we are able for the first
time to fill in the gaps in the phase diagram, which remain elusive in previous works.

Figure~\ref{phase-diagram} shows the phase diagram of the CO$_2$--water mixture. The thermodynamic conditions corresponding to the H--L$_{\text{H}_2\text{O}}$--V and H--L$_{\text{H}_2\text{O}}$--L$_{\text{CO}_2}$ three-phase equilibria are listed in Table~\ref{P-T3}. The equilibrium line between the hydrate, liquid water, and liquid CO$_2$ phases was accurately reproduced, including the reentrant behavior, in previous MD simulations (filled-cyan circles) conducted by some of us\cite{Miguez2015a} using the traditional three-phase direct coexistence technique. However, accessing the equilibrium line with the vapor phase (H-L$_{\text{H}_2\text{O}}$-V) was previously challenging.
We now plot the calculations performed in this work (filled-red circles) for the equilibrium line with the vapor phase, using the solubility method, and remarkably, it aligns with the experimental line. This exciting development now allows us to calculate the quadruple point Q$_{2}$ as the intersection between our new calculations on the H-L$_{\text{H}_2\text{O}}$-V coexistence and those on the H-L$_{\text{H}_2\text{O}}$-L$_{\text{CO}_{2}}$ coexistence.
Our estimation of the Q$_{2}$ point (indicated by a violet circle in Fig.~\ref{phase-diagram}) is $283.5\,\text{K}$ and $44\,\text{bar}$. Notably, this value agrees remarkably well (within the uncertainty) with the Q$_2$ measurement, which is reported as $283\,\text{K}$ and $45\,\text{bar}$.\cite{Sloan2008a} This strong agreement between our estimated quadruple point and experimental data reinforces the reliability of our methodology and the accuracy of the force field employed in this work.

 \begin{table}
\caption{Three-phase coexistence temperatures, $T_3$, of CO$_{2}$ hydrate obtained from molecular simulations at different pressures $P$. Results for 10–$40\,\text{bar}$ (H-L$_{\text{H}_2\text{O}}$-V) are obtained in this work, whereas data for 50–$4000\,\text{bar}$ (H-L$_{\text{H}_2\text{O}}$-L$_{\text{CO}_{2}}$) were reported previously.~\cite{Miguez2015a} $T_3^{exp}$ denotes the corresponding experimental values taken from the literature.~\cite{Sloan2008a} The last column lists the phases coexisting under the thermodynamic conditions of $P$ and $T_3$.}
\label{P-T3}
\centering
\begin{tabular}{cccc}
\hline\hline
$P$ (bar) & $T_3$ (K) & $T_3^{exp}$ (K) & Phases    \\
\hline
10 & 271(1) & - & H-L$_{\text{H}_2\text{O}}$-V$^*$ \\
15 & 273.3(8) & 274.8 & H-L$_{\text{H}_2\text{O}}$-V \\
20 & 276.8(8) & 277.6 & H-L$_{\text{H}_2\text{O}}$-V \\
30 & 279.6(7) & 280.3 & H-L$_{\text{H}_2\text{O}}$-V \\
40 & 282.4(8)& 282.4 & H-L$_{\text{H}_2\text{O}}$-V \\
\hline
50 & 284(2) & 282.9 & H-L$_{\text{H}_2\text{O}}$-L$_{\text{CO}_{2}}$ \\
100 & 284(2) & 283.6 & H-L$_{\text{H}_2\text{O}}$-L$_{\text{CO}_{2}}$ \\
200 & 284(2) & 284.6 & H-L$_{\text{H}_2\text{O}}$-L$_{\text{CO}_{2}}$ \\
400 & 287(2) & 286.2 & H-L$_{\text{H}_2\text{O}}$-L$_{\text{CO}_{2}}$ \\
1000 &289(2) & 289.7 & H-L$_{\text{H}_2\text{O}}$-L$_{\text{CO}_{2}}$ \\
2000 & 292(2) & 293.0 & H-L$_{\text{H}_2\text{O}}$-L$_{\text{CO}_{2}}$ \\
3000 & 287(2) & 293.9 & H-L$_{\text{H}_2\text{O}}$-L$_{\text{CO}_{2}}$ \\
4000 & 284(2) & 293.6 & H-L$_{\text{H}_2\text{O}}$-L$_{\text{CO}_{2}}$ \\
\hline
\hline
\end{tabular}

\justifying
\noindent
\scriptsize $^*$Notice that the simulation result obtained at $10\,\text{bar}$ corresponds to a metastable point and, hence, there is no experimental data for the H-L$_{\text{H}_2\text{O}}$-V equilibria at such pressure. Experimentally,~\cite{Sloan2008a} at  $12.56\,\text{bar}$ there exists a quadrupole point (Q$_1$) where the CO$_2$ hydrate, an ice Ih, a liquid rich-water, and a vapor phases coexist. Under the Q$_1$ pressure, the liquid rich-water phase becomes unstable and only three phases should coexist: CO$_2$ hydrate, ice Ih, and vapor. 
\end{table}

\begin{figure}
\includegraphics[width=\columnwidth]{phase-diagram-co2_V3.eps}
\caption{Phase diagram of the CO$_{2}$ + water mixture assuming an excess of CO$_2$. The open black symbols represent the experimental three-phase lines for the H-L$_{\text{H}_2\text{O}}$-L$_{\text{CO}_{2}}$ (diamonds),~\cite{Sloan2008a} H-L$_{\text{H}_2\text{O}}$-V (squares),~\cite{Sloan2008a} H-L$_{\text{CO}_{2}}$-V (left triangles),~\cite{Sloan2008a} L$_{\text{H}_2\text{O}}$-L$_{\text{CO}_{2}}$-V (right triangles),~\cite{Sloan2008a} and H-Ih-V (down triangles)~\cite{Sloan2008a} equilibria. The filled-black square and circle are the experimental quadruple points Q$_{1}$ and Q$_{2}$,~\cite{Sloan2008a} respectively.  Cyan-filled circles correspond to simulation results obtained by some of us in a previous work~\cite{Miguez2015a} for the H-L$_{\text{H}_2\text{O}}$-L$_{\text{CO}_{2}}$ three-phase line. Red-filled circles correspond to the three-phase line for the H-L$_{\text{H}_2\text{O}}$-V equilibrium obtained in this work. The filled violet circle is the simulation quadruple point Q$_2$ obtained in this work.}
\label{phase-diagram}
\end{figure}

\section{Conclusions}

In this work, for the first time, computer simulations have successfully determined the three-phase coexistence equilibrium conditions of the CO$_2$ and water mixture. This achievement has been possible by developing a novel methodology based on solubility calculations and combined with previous DC simulations of condensed phases.
Moreover, the successful alignment of our computed equilibrium lines with experimental data for both the H-L$_{\text{H}_2\text{O}}$-L$_{\text{CO}_{2}}$ coexistence and the H-L$_{\text{H}_2\text{O}}$-V coexistence enables us to confidently estimate an upper quadruple point Q$_2$ which is in excellent agreement with experimental observations. Our findings demonstrate the significant advancements made in understanding the CO$_2$ and water mixture's phase behavior and provide valuable insights into the behavior of gas hydrates at different conditions. This methodology is expected to be applicable in the vapor region for other hydrates that exhibit an H-L$_{\text{H}_2\text{O}}$-V three-phase line, thereby paving the way for the systematic determination of additional hydrate dissociation lines. Such an extension would represent a significant step forward toward a more comprehensive understanding of hydrate phase behavior under low-pressure conditions.

\section*{Acknowledgements}
CR-G, JA, and FJB acknowledge grant Refs.~PID2021-125081NB-I00 and PID2024-158030NB-I00 funded both by MCIN/AEI/10.13039/501100011033 and FEDER EU, and Universidad de Huelva (P.O. FEDER EPIT1282023), also co-funded by EU FEDER funds. MMC also acknowledges grant Ref.~PID2022-136919NB-C32 funded by MCIN/AEI/10.13039/501100011033. CR-G acknowledges the FPI Grant (Ref.~PRE2022-102927) from Ministerio de Ciencia e Innovación and Fondo Social Europeo Plus. CV also acknowledges grant Ref.~PID2022-136919NB-C31 funded by MCIN/AEI/10.13039/501100011033. We also greatly acknowledge RES resources provided by the Bioinnovation Center of the University of Malaga in Picasso to FI-2024-2-0030.  The authors gratefully acknowledge the Universidad Politecnica de Madrid (www.upm.es) for providing computing resources on Magerit Supercomputer. The authors would like to express our sincere gratitude to Carlos Vega in the Special Issue published in his honor. Working with him is truly a privilege, not only because of everything one learns, but also because even when you think you understand something perfectly, Carlos can ask a single question that makes you rethink everything … and somehow, he is always right. We look forward to continuing these discussions soon, ideally over a {\it Cocido Madrileño} and a {\it Vermut} at the Taberna de Ángel Sierra.

\section*{Conflicts of interest}

The authors have no conflicts to disclose.

\section*{Author contributions}

\noindent
\textbf{Jesús Algaba:} Conceptualization (equal); Methodology (equal); Investigation (equal); Writing – original draft (equal); Writing – review \& editing (equal). \textbf{Samuel Blazquez:} Methodology (equal); Investigation (equal); Writing – original draft (equal); Writing – review \& editing (equal). \textbf{Cristóbal Romero-Guzmán:} Methodology (equal); Writing – review \& editing (equal).
\textbf{Carlos Vega:} Funding acquisition (equal); Methodology (equal); Writing – review \& editing (equal). \textbf{María M. Conde:} Conceptualization (equal); Funding acquisition (equal); Methodology (equal); Writing – review \& editing (equal). \textbf{Felipe J. Blas:} Conceptualization (equal); Funding acquisition (equal); Methodology (equal); Writing – review \& editing (equal).

\section*{Data availability}

The data that support the findings of this study are available within the article.

\section*{References}
\bibliography{masterbib}

\end{document}